\documentclass[zpreprint,zbstpl]{zeus_paper}

\usepackage[english]{babel}

\chardef\usc=95
\chardef\til=126
\catcode`\@=11 
\DeclareRobustCommand\xdotspace{\futurelet\@let@token\@xdotspace}
\def\@xdotspace{%
  \ifx\@let@token.\else
  \ifx\@let@token\bgroup.\else
  \ifx\@let@token\egroup.\else
  \ifx\@let@token\/.\else
  \ifx\@let@token\ .\else
  \ifx\@let@token~.\else
  \ifx\@let@token!.\else
  \ifx\@let@token,.\else
  \ifx\@let@token:.\else
  \ifx\@let@token;.\else
  \ifx\@let@token?.\else
  \ifx\@let@token/.\else
  \ifx\@let@token'.\else
  \ifx\@let@token).\else
  \ifx\@let@token-.\else
  \ifx\@let@token\@xobeysp.\else
  \ifx\@let@token\space.\else
  \ifx\@let@token\@sptoken.\else
   .\space
   \fi\fi\fi\fi\fi\fi\fi\fi\fi\fi\fi\fi\fi\fi\fi\fi\fi\fi}
\catcode`\@=12 

\newcommand{\stru}[2]{%
   \relax\ifmmode\hbox{\vrule height#1 depth#2 width0pt}%
   \else\vrule height#1 depth#2 width0pt\fi}

\newcommand{\Ronum}[1]{\uppercase\expandafter{\romannumeral#1}}
\newcommand{\ronum}[1]{\expandafter{\romannumeral#1}}
\DeclareRobustCommand{\LaTeXZ}{%
  \LaTeX\kern-.05em4\kern-.1em
  {\raisebox{-0.2ex}{$\scriptstyle\text{ZEUS}$}}\xspace}

\DeclareMathAlphabet{\mathbf}{OT1}{cmr}{bx}{sl}
\newcommand{\eVdist}{\kern-0.06667em}

\newcommand{\Gev}{{\text{Ge}\eVdist\text{V\/}}}

\newcommand{\gev}{{\,\text{Ge}\eVdist\text{V\/}}}

\newcommand{\pb}{\,\text{pb}}
\newcommand{\fb}{\,\text{fb}}

\newcommand{\Tesla}{\,\text{T}}

\newcommand{\slashfrac}[2]{%
  \raisebox{0.5ex}{\ensuremath #1}\kern-0.12em/\kern-0.08em
  \raisebox{-.8ex}{\ensuremath #2}}

\newcommand{\sqr}[3]{%
    {\vcenter{\hrule height.#3ex\hbox{\vrule width.#2ex height#1ex
     \kern#1ex\vrule width.#3ex}\hrule height.#2ex}}}

\catcode`\@=11 
\newcommand{\parenbar}{\mathpalette\p@renb@r}
\def\p@renb@r#1#2{\vbox{%
  \ifx#1\scriptscriptstyle \dimen@.7em\dimen@ii.2em\else
  \ifx#1\scriptstyle \dimen@.8em\dimen@ii.25em\else
  \dimen@1em\dimen@ii.4em\fi\fi \offinterlineskip
  \ialign{\hfill##\hfill\cr
    \vbox{\hrule width\dimen@ii}\cr
    \noalign{\vskip-.3ex}%
    \hbox to\dimen@{$\mathchar300\hfil\mathchar301$}\cr
    \noalign{\vskip-.3ex}%
    $#1#2$\cr}}}
\catcode`\@=12 

\newcommand{\IP}{{\rm I$\kern-0.01667em$P}\xspace}

\mathchardef\qsm=63
\mathchardef\pls=43
\mathchardef\mns=512
\mathchardef\plm=518
\mathchardef\eql=61
\mathchardef\smallleft=300
\mathchardef\smallright=301
\mathchardef\les=316
\mathchardef\gre=318
\mathchardef\leq=532
\mathchardef\grq=533

\catcode`\@=11 
\newcounter{pict@width}
\newcounter{pict@height}
\newlength{\pict@scale}
\newcommand{\psfigadd}[4]{%
\setcounter{pict@width}{1*\ratio{#2+\pict@scale/2}{\pict@scale}}
\setcounter{pict@height}{1*\ratio{#3+\pict@scale/2}{\pict@scale}}
\setlength{\unitlength}{\pict@scale}
\hbox to #2{\hspace{-\fill}\begin{picture}(\thepict@width,\thepict@height)
\put(0,0){\psfig{figure=#1,width=#2,height=#3,clip=}}
\SetScale{0.283466457}
\SetWidth{1.763889}
{#4}
\end{picture}}
}
\newcounter{pict@widthfst}
\newcounter{pict@widthscd}
\newcounter{pict@widthtot}
\newcommand{\psfigaddtwo}[7]{%
\setcounter{pict@widthfst}{1*\ratio{#2+\pict@scale/2}{\pict@scale}}
\setcounter{pict@widthscd}{1*\ratio{#2+#4+\pict@scale/2}{\pict@scale}}
\setcounter{pict@widthtot}{1*\ratio{#2+#4+#6+\pict@scale/2}{\pict@scale}}
\setcounter{pict@height}{1*\ratio{#3+\pict@scale/2}{\pict@scale}}
\setlength{\unitlength}{\pict@scale}
\hbox{\hspace{-\fill}\begin{picture}(\thepict@widthtot,\thepict@height)
\put(0,0){\psfig{figure=#1,width=#2,height=#3,clip=}}
\put(\thepict@widthscd,0){\psfig{figure=#5,width=#6,height=#3,clip=}}
\SetScale{0.283466457}
\SetWidth{1.763889}
{#7}
\end{picture}}
}
\newcommand{\psfigror}[4]{%
\setcounter{pict@width}{1*\ratio{#2+\pict@scale/2}{\pict@scale}}
\setcounter{pict@height}{1*\ratio{#3+\pict@scale/2}{\pict@scale}}
\setlength{\unitlength}{\pict@scale}
\hbox{\begin{picture}(\thepict@width,\thepict@height)
\put(0,\thepict@height){\psfig{figure=#1,width=#3,height=#2,clip=,angle=270}}
\SetScale{0.283466457}
\SetWidth{1.763889}
{#4}
\end{picture}}
}
\newcommand{\psfigrol}[4]{%
\setcounter{pict@width}{1*\ratio{#2+\pict@scale/2}{\pict@scale}}
\setcounter{pict@height}{1*\ratio{#3+\pict@scale/2}{\pict@scale}}
\setlength{\unitlength}{\pict@scale}
\hbox{\begin{picture}(\thepict@width,\thepict@height)
\put(0,0){\psfig{figure=#1,width=#3,height=#2,clip=,angle=90}}
\SetScale{0.283466457}
\SetWidth{1.763889}
{#4}
\end{picture}}
}
\catcode`\@=12 
\newlength\listtextwidth

\catcode`\@=11 
\newlength{\@tabfninsert}
\newlength{\@tabfnwidth}
\newcommand{\tabfootnote}[2]{%
  \setlength{\@tabfninsert}{0.8em}
  \setlength{\@tabfnwidth}{\textwidth}
  \addtolength{\@tabfnwidth}{-\@tabfninsert}
  \addtolength{\@tabfnwidth}{-0.4em}
  \noindent\makebox[\@tabfninsert][r]{\footnotesize$^{#1}$\hfil}\hfill%
  \parbox[t]{\@tabfnwidth}{\footnotesize #2\hfill}}
\catcode`\@=12 

\def\etjet{E_T^{\rm jet}}

\def\etaj{\eta^{\rm jet1}}

\def\etj{E_T^{\rm jet1}}

\def\phij{\varphi^{\rm jet1}}

\def\setjb{d\sigma/d\etjb}
\def\sq2{d\sigma/d\q2}
 
\def\q2{Q^2}

\def\cgh{\cos\gamma_h}
\def\pb1{pb$^{-1}$}
\def\fb1{fb$^{-1}$}

\def\g2{GeV$^2$}

\def\rr1{R=1}
\def\r7{R=0.7}
\def\R71{R=0.7\ {\rm and}\ 1}
\def\RRR{R=1,\ 0.7\ {\rm and}\ 0.5}

\def\kt{k_T}

\def\lq2{\log_{10}(\q2)}

\def\etjb{E^{\rm jet}_{T,{\rm B}}}

\def\etajb{\eta^{\rm jet}_{\rm B}}

\def\etalab{\eta^{\rm jet}_{\rm LAB}}
\def\etlab{E^{\rm jet}_{T,{\rm LAB}}}

\def\colab#1{#1 Coll.}

\def\as{\alpha_s}
\def\oalphas2{{\cal O}(\alpha\as^2)}

\def\oas{{\cal O}(\as)}
\def\oass{{\cal O}(\as^2)}

\def\asz{\as(\mz)}
\def\asmz#1#2#3#4#5#6{\asz = #1\pm #2\ {\rm (stat.)}\ ^{+#4}_{-#3}\ {\rm (exp.)}\ ^{+#6}_{-#5}\ {\rm (th.)}}
\def\mz{M_Z}
\def\qq{q\bar q}
\def\z0{Z^0}
\def\ele{e^+e^-}
\def\JHEP{JHEP}

\def\etal{et al.}

\def\etai{\eta^i_{\rm B}}
\def\etaj{\eta^j_{\rm B}}
\def\eti{E^i_{T,{\rm B}}}
\def\etj{E^j_{T,{\rm B}}}
\def\phii{\phi^i_{\rm B}}
\def\phij{\phi^j_{\rm B}}

\def\figdir{./}

\begin{document}
\prepnum{{DESY--06--241}}

\title{
         Jet-radius dependence 
of inclusive-jet cross sections in deep inelastic scattering at HERA
}                                                       
                    
\author{ZEUS Collaboration}
\date{December 2006}

\abstract{
Differential inclusive-jet cross sections have been measured for
different jet radii in neutral current deep inelastic $ep$ scattering
for boson virtualities $\q2>125$~\g2\ with the ZEUS detector at
HERA using an integrated luminosity of $81.7$~\pb1. Jets were
identified in the Breit frame using the $\kt$ cluster algorithm in the
longitudinally inclusive mode for different values of the jet radius
$R$. Differential cross sections are presented as functions of $\q2$
and the jet transverse energy, $\etjb$. The dependence on $R$ of the
inclusive-jet cross section has been measured for $\q2>125$ and
$500$~\g2\ and found to be linear with $R$ in the range
studied. Next-to-leading-order QCD calculations give a good
description of the measurements for $0.5\leq R\leq 1$. A value of
$\asz$ has been extracted from the measurements of the inclusive-jet
cross-section $\sq2$ with $R=1$ for $\q2>500$~\g2:
$\asmz{0.1207}{0.0014}{0.0033}{0.0035}{0.0023}{0.0022}$. The
variation of $\as$ with $\etjb$ is in good agreement with the running
of $\as$ as predicted by QCD.
}

\makezeustitle

\def\3{\ss}

\pagenumbering{Roman}

\begin{center}
{                      \Large  The ZEUS Collaboration              }
\end{center}

  S.~Chekanov$^{   1}$,
  M.~Derrick,
  S.~Magill,
  S.~Miglioranzi$^{   2}$,
  B.~Musgrave,
  D.~Nicholass$^{   2}$,
  \mbox{J.~Repond},
  R.~Yoshida\\
 {\it Argonne National Laboratory, Argonne, Illinois 60439-4815},
        USA~$^{n}$
\par \filbreak

  M.C.K.~Mattingly \\
 {\it Andrews University, Berrien Springs, Michigan 49104-0380}, USA
\par \filbreak

  M.~Jechow, N.~Pavel~$^{\dagger}$, A.G.~Yag\"ues Molina \\
  {\it Institut f\"ur Physik der Humboldt-Universit\"at zu Berlin,
        Berlin, Germany}
% ...

\newpage

$^{\    1}$ supported by DESY, Germany \\
$^{\    2}$ also affiliated with University College London, UK \\
% ...
$^{\dagger}$ deceased \\

\newpage

\begin{tabular}[h]{rp{14cm}}

$^{a}$ &  supported by the Natural Sciences and Engineering Research
Council of Canada (NSERC) \\
$^{b}$ &  supported by the German Federal Ministry for Education and
Research (BMBF), under contract numbers HZ1GUA 2, HZ1GUB 0, HZ1PDA 5,
HZ1VFA 5\\
$^{c}$ &  supported in part by the MINERVA Gesellschaft f\"ur
Forschung GmbH, the Israel Science Foundation (grant no. 293/02-11.2)
and the U.S.-Israel Binational Science Foundation \\
$^{d}$ &  supported by the German-Israeli Foundation and the Israel
Science Foundation\\
$^{e}$ &  supported by the Italian National Institute for Nuclear
Physics (INFN) \\
$^{f}$ &  supported by the Japanese Ministry of Education, Culture,
Sports, Science and Technology (MEXT) and its grants for Scientific
Research\\
$^{g}$ &  supported by the Korean Ministry of Education and Korea
Science and Engineering Foundation\\
$^{h}$ &  supported by the Netherlands Foundation for Research on
Matter (FOM)\\
$^{i}$ &  supported by the Polish State Committee for Scientific
Research, grant no. 620/E-77/SPB/DESY/P-03/DZ 117/2003-2005 and grant
no. 1P03B07427/2004-2006\\
$^{j}$ &  partially supported by the German Federal Ministry for
Education and Research (BMBF)\\
$^{k}$ &  supported by RF Presidential grant N 1685.2003.2 for the
leading scientific schools and by the Russian Ministry of Education
and Science through its grant for Scientific Research on High Energy
Physics\\
$^{l}$ &  supported by the Spanish Ministry of Education and Science
through funds provided by CICYT\\
$^{m}$ &  supported by the Particle Physics and Astronomy Research
Council, UK\\
$^{n}$ &  supported by the US Department of Energy\\
$^{o}$ &  supported by the US National Science Foundation. Any
opinion, findings and conclusions or recommendations expressed in this
material are those of the authors and do not necessarily reflect the
views of the National Science Foundation.\\
$^{p}$ &  supported by the Polish Ministry of Science and Higher
Education\\
$^{q}$ &  supported by FNRS and its associated funds (IISN and FRIA)
and by an Inter-University Attraction Poles Programme subsidised by
the Belgian Federal Science Policy Office\\
$^{r}$ &  supported by the Malaysian Ministry of Science, Technology
and Innovation/Akademi Sains Malaysia grant SAGA 66-02-03-0048\\

\end{tabular}

\newpage

\pagenumbering{arabic} 
\pagestyle{plain}

\pagenumbering{arabic} 
\pagestyle{plain}

\section{Introduction}
\label{intro}

The study of jet production in $ep$ collisions at HERA has been well
established as a testing ground of perturbative QCD (pQCD) providing
precise determinations of the strong coupling constant, $\as$, 
and its scale dependence. The jet observables used to test pQCD
included dijet~\cite{pl:b507:70,epj:c19:289,epj:c23:13,hep-ex-0608048},
inclusive-jet~\cite{epj:c19:289,pl:b547:164,pl:b551:226,hep-ex-0608048}
and multijet~\cite{pl:b515:17,epj:c44:183} cross sections in neutral
current (NC) 
deep inelastic $ep$ scattering (DIS),
dijet~\cite{epj:c11:35,epj:c23:615,pl:b531:9,epj:c25:13,pl:b639:21},
inclusive-jet~\cite{pl:b560:7,epj:c29:497} and
multijet~\cite{pl:b443:394} cross sections in photoproduction and the
internal structure of jets in
NC~\cite{np:b545:3,pl:b558:41,np:b700:3} and charged
current~\cite{epj:c31:149} DIS.

These studies demonstrated that the $\kt$ cluster 
algorithm~\cite{np:b406:187} in the longitudinally invariant inclusive 
mode~\cite{pr:d48:3160} is at present the method to reconstruct jets in 
$ep$ collisions for which the smallest uncertainties are achieved.
Previous analyses were done with jet radius $R=1$, where $R$
is the maximum distance in the pseudorapidity ($\eta$) - azimuth ($\phi$)
plane for particle recombination. The study of the predicted jet cross
sections using different jet radii allows the identification of the
values of $R$ for which the theory is most reliable. Furthermore, 
smaller values of the jet radius $R$ are of particular interest for
the identification of heavy particles decaying into jets; in these
decays, the final-state jets may emerge close in phase space and need
to be identified separately for a faithful reconstruction of the
properties of the parent particle~\cite{zfp:c62:127}. Neutral current
DIS provides a well understood environment in which to study the
dependence of jet production on the jet radius and to confront the
data with precise next-to-leading-order (NLO) QCD calculations in a
hadron-induced reaction.

The hadronic final state in NC DIS may consist of jets of high
transverse energy, $\etjet$, produced in the hard-scattering process. 
In NC DIS, the Breit frame~\cite{bookfeynam:1972,*zfp:c2:237} is
preferred to conduct the jet search, since jet production is directly
sensitive to hard QCD processes; in the Born process 
($eq\rightarrow eq$), the virtual boson ($V^{*}$, with $V^{*}=\gamma, Z$)
is absorbed by the struck quark, which is back-scattered with zero 
transverse momentum with respect to the $V^{*}$ direction. At leading
order (LO) in $\as$, the boson-gluon-fusion ($V^{*} g\rightarrow \qq$)  
and QCD-Compton ($V^{*} q \rightarrow qg$) processes give rise to two
hard jets with opposite transverse momenta. Inclusive-jet production
at high $\etjet$ allows more stringent tests of the pQCD calculations
than dijet production due to the reduced theoretical uncertainties. 
Restrictions on the topology of dijet events are necessary to avoid
infrared-sensitive regions where the NLO QCD programs are not
reliable, whereas inclusive-jet cross sections are infrared insensitive. 
Therefore, such measurements allow tests of pQCD in the widest
phase-space region for jet production. In particular, previous
measurements of inclusive-jet cross sections in NC DIS at high
$\q2$~\cite{pl:b547:164}, where $\q2$ is the negative of the square of
the four-momentum transfer, provided the most precise determination
of $\as$ at HERA to date.

This letter presents new measurements of differential inclusive-jet
cross sections as a function of the jet transverse energy in the Breit
frame, $\etjb$, and $\q2$ for different values of $R$. For $R=1$, this
analysis is based on the same inclusive-jet data sample presented in a
recent publication~\cite{hep-ex-0608048}. The results have been
compared with NLO QCD calculations using recent parameterisations of
the parton distribution functions (PDFs) of the
proton~\cite{pr:d67:012007,epj:c28:455,jhep:0207:012,*jhep:0310:046}.
In addition, an updated determination of $\as$ and of its scale
dependence has been obtained using a data sample which corresponds to
more than a twofold increase in luminosity with respect to the previous
analysis~\cite{pl:b547:164}.

\section{Experimental set-up}
A detailed description of the ZEUS detector can be found
elsewhere~\cite{pl:b293:465,zeus:1993:bluebook}. A brief outline of
the components that are most relevant for this analysis is given below.

Charged particles are tracked in the central tracking detector
(CTD)~\cite{nim:a279:290,*npps:b32:181,*nim:a338:254}, which operates
in a magnetic field of $1.43\Tesla$ provided by a thin superconducting
solenoid. The CTD consists of $72$~cylindrical drift-chamber
layers, organised in nine superlayers covering the
polar-angle\footnote{The ZEUS coordinate system is a right-handed
  Cartesian system, with the $Z$ axis pointing in the proton beam
  direction, referred to as the ``forward direction'', and the $X$
  axis pointing left towards the centre of HERA. The coordinate origin
  is at the nominal interaction point.}
region \mbox{$15^\circ<\theta<164^\circ$}. The transverse-momentum
resolution for full-length tracks can be parameterised as
$\sigma(p_T)/p_T=0.0058p_T\oplus0.0065\oplus0.0014/p_T$, with $p_T$ in
$\Gev$. The tracking system was used to measure the interaction vertex
with a typical resolution along (transverse to) the beam direction of
$0.4$~($0.1$)~cm and to cross-check the energy scale of the calorimeter.

The high-resolution uranium--scintillator calorimeter
(CAL)~\cite{nim:a309:77,*nim:a309:101,*nim:a321:356,*nim:a336:23} covers 
$99.7\%$ of the total solid angle and consists of three parts: the
forward (FCAL), the barrel (BCAL) and the rear (RCAL) calorimeters. 
Each part is subdivided transversely into towers and longitudinally
into one electromagnetic section and either one (in RCAL) or two (in
BCAL and FCAL) hadronic sections. The smallest subdivision of the
calorimeter is called a cell. Under test-beam conditions, the CAL
single-particle relative energy resolutions were
$\sigma(E)/E=0.18/\sqrt E$ for electrons and 
$\sigma(E)/E=0.35/\sqrt E$ for hadrons, with $E$ in GeV.

The luminosity was measured from the rate of the bremsstrahlung process
$ep\rightarrow e\gamma p$. The resulting small-angle energetic photons
were measured by the luminosity
monitor~\cite{desy-92-066,*zfp:c63:391,*acpp:b32:2025}, a
lead-scintillator calorimeter placed in the HERA tunnel at $Z=-107$ m.

\section{Data selection and jet search}
The data were collected during the running period 1998-2000, when HERA
operated with protons of energy $E_p=920$~GeV and electrons or
positrons\footnote{Here and in the following, the term ``electron''
  denotes generically both the electron ($e^-$) and the positron
  ($e^+$).} of energy $E_e=27.5$~GeV, and
correspond to an integrated luminosity of $81.7\pm 1.8$~\pb1, of which
$16.7$~\pb1\ ($65.0$~\pb1) was for $e^-p$ ($e^+p$) collisions.

Neutral current DIS events were selected offline using the same
criteria as reported in a recent
publication~\cite{hep-ex-0608048}. The main steps are briefly listed
below.
 
The scattered-electron candidate was identified from the pattern of
energy deposits in the CAL~\cite{nim:a365:508,*nim:a391:360}. The
energy ($E_e^{\prime}$) and polar angle ($\theta_e$) of the
electron candidate were determined from the CAL measurements. 
The $\q2$ variable was reconstructed using the double angle method
($\q2_{\rm DA}$)~\cite{proc:hera:1991:23,*proc:hera:1991:43}.
The angle $\gamma_h$, which corresponds to the angle of the scattered
quark in the quark-parton model, was reconstructed using the hadronic
final state~\cite{proc:hera:1991:23,*proc:hera:1991:43}.

The main requirements imposed on the data sample were:
an electron candidate with $E_{e}^{\prime}>10$~GeV;
a vertex position along the beam axis in the range $|Z|<34$~cm; 
$38<(E-P_Z)<65$~GeV, where $E$ is the total energy as measured
by the CAL, $E=\sum_iE_i$, and $P_Z$ is the $Z$-component of the
vector ${\bf P}=\sum_i {E_i} \bf{r_i}$; in both cases the sum runs
over all CAL cells, $E_i$ is the energy of the CAL cell $i$
and ${\bf r_i}$ is a unit vector along the line joining the
reconstructed vertex and the geometric centre of the cell $i$;
$\q2>125$~\g2; and $|\cgh|<0.65$. After all these requirements,
contamination from non-$ep$ interactions and other physics processes
was negligible.

The $\kt$ cluster algorithm was used in the longitudinally invariant
inclusive mode to reconstruct jets in the hadronic final state both in
data and in Monte Carlo (MC) simulated events (see Section~\ref{mc}). 
In the data, the algorithm was applied to the energy deposits measured in
the CAL cells after excluding those associated with the
scattered-electron candidate. In the following discussion, $\eti$
denotes the transverse energy, $\etai$ the pseudorapidity and
$\phii$ the azimuthal angle of object $i$ in the Breit frame. For
each pair of objects, where the initial objects are the energy
deposits in the CAL cells, the quantity
$$d_{ij}={\rm min}(\eti,\etj)^2\cdot[(\etai-\etaj)^2+(\phii-\phij)^2]/R^2$$
was calculated. For each individual object, the quantity $d_i=(\eti)^2$
was also calculated. If, of all the values $\{d_{ij},d_i\}$, $d_{kl}$ was
the smallest, then objects $k$ and $l$ were combined into a single new
object. If, however, $d_k$ was the smallest, then object $k$ was 
considered a jet and was removed from the sample. The procedure was
repeated until all objects were assigned to jets. The jet search was
performed with different jet radii ($R=0.5$, $0.7$ and $1$) and the
jet variables were defined according to the Snowmass
convention~\cite{proc:snowmass:1990:134}.

After reconstructing the jet variables in the Breit frame, the
massless four-momenta were boosted into the laboratory frame, where
the transverse energy~($\etlab$) and the pseudorapidity~($\etalab$) of
each jet were calculated. Energy
corrections~\cite{pl:b547:164,pl:b531:9,pl:b558:41} were then applied
to the jets in the laboratory frame and propagated back into $\etjb$. The
jet variables in the laboratory frame were also used to apply
additional cuts on the selected sample, as explained in a recent
publication~\cite{hep-ex-0608048}: events were removed from the sample
if any of the jets was in the backward region of the detector
($\etalab<-2$) and jets were not included in the final sample if
$\etlab<2.5$~GeV.

Only events with at least one jet in the pseudorapidity range
$-2<\etajb<1.5$ were kept for further analysis. The final data samples
with at least one jet satisfying $\etjb>8$~GeV contained $19908$
events for $R=1$, $16231$ for $R=0.7$ and $12935$ for $R=0.5$.

\section{Monte Carlo simulation}
\label{mc}
Samples of events were generated to determine the response of the
detector to jets of hadrons and the correction factors necessary to
obtain the hadron-level jet cross sections. The hadron level is
defined by those hadrons with lifetime $\tau\geq 10$~ps. The generated
events were passed through the {\sc
  Geant}~3.13-based~\cite{tech:cern-dd-ee-84-1} ZEUS detector- and
trigger-simulation programs~\cite{zeus:1993:bluebook}. They were
reconstructed and analysed by the same program chain as the data.
 
Neutral current DIS events including electroweak radiative effects
were simulated using the 
{\sc Heracles}~4.6.1~\cite{cpc:69:155,*spi:www:heracles} program with
the {\sc Djangoh}~1.1~\cite{cpc:81:381,*spi:www:djangoh11} interface
to the QCD programs. The QCD cascade is simulated using the
colour-dipole model
(CDM)~\cite{pl:b165:147,*pl:b175:453,*np:b306:746,*zfp:c43:625} 
including the leading-order (LO) QCD diagrams as implemented in
{\sc Ariadne}~4.08~\cite{cpc:71:15,*zfp:c65:285} and, alternatively, 
with the MEPS model of {\sc Lepto} 6.5~\cite{cpc:101:108}. 
The CTEQ5D~\cite{epj:c12:375} parameterisations of the proton PDFs were
used for these simulations. Fragmentation into hadrons is performed
using the Lund string model~\cite{prep:97:31} as implemented in 
{\sc Jetset}~\cite{cpc:82:74,*cpc:135:238,cpc:39:347,*cpc:43:367}.
 
The jet search was performed on the MC events using the energy
measured in the CAL cells in the same way as for the data. 
The same jet algorithm was also applied to the final-state particles
(hadron level) and to the partons available after the parton shower
(parton level). The MC programs were also used to correct the measured
cross sections for QED radiative effects and the running of 
$\alpha_{\rm em}$.

\section{QCD calculations}
\label{nlo}
The measurements were compared with LO ($\oas$) and NLO QCD ($\oass$)
calculations obtained using the program {\sc Disent}~\cite{np:b485:291}. 
The calculations were performed in the $\overline{\rm MS}$ renormalisation
and factorisation schemes using a generalised
version~\cite{np:b485:291} of the subtraction 
method~\cite{np:b178:421}. The number of flavours was set to five and
the renormalisation ($\mu_R$) and factorisation ($\mu_F$) scales were
chosen to be $\mu_R=\etjb$ and $\mu_F=Q$, respectively. The strong
coupling constant was calculated at two loops with
$\Lambda^{(5)}_{\overline{\rm MS}}=226$~MeV, corresponding to
$\asz=0.118$. The calculations were performed using the
ZEUS-S~\cite{pr:d67:012007} parameterisations of the proton PDFs. In
{\sc Disent}, the value of $\alpha_{\rm em}$ was fixed to $1/137$. The
$\kt$ cluster algorithm was also applied to the partons in the events
generated by {\sc Disent} in order to compute the jet cross-section
predictions. The calculations were performed for the same values of
$R$ as for the data. In addition, predictions were obtained for
$R=0.3$ and $1.2$ to determine the range of $R$ in which the theory is
most reliable.
 
Since the measurements refer to jets of hadrons, whereas the NLO QCD
calculations refer to jets of partons, the predictions were corrected
to the hadron level using the MC models. The multiplicative correction
factor ($C_{\rm had}$) was defined as the ratio of the cross section
for jets of hadrons over that for jets of partons, estimated by using
the MC programs described in Section~\ref{mc}. The mean of the ratios 
obtained with {\sc Ariadne} and {\sc Lepto-MEPS} was taken as the value
of $C_{\rm had}$. The value of $C_{\rm had}$ differs from unity by 
less than $5\%$, $15\%$ and $25\%$ for $R=1$, $0.7$ and $0.5$, 
respectively, in the region $\q2\geq 500$~\g2. For $R=1.2$, 
$C_{\rm had}$ differs from unity by less than $1\%$ and for $R=0.3$,
$C_{\rm had}$ differs by $40\%$. 

{\sc Disent} does not include the contribution from $\z0$ exchange; MC
simulated events with and without $\z0$ exchange were used to include
this effect in the NLO QCD predictions. In the following, NLO QCD
calculations will refer to the fully corrected predictions, except
where otherwise stated.

\subsection{Theoretical uncertainties}
\label{thun}
Several sources of uncertainty in the theoretical predictions were
considered:
\begin{itemize}
  \item the uncertainty on the NLO QCD calculations due to terms
    beyond NLO, estimated by varying $\mu_R$ between $\etjb/2$ and
    $2\etjb$, was below $\pm 7\%$ at low $\q2$ and low $\etjb$ and
    decreased to less than $\pm 5\%$ for $\q2 > 250$~\g2\ for $R=1$.
    For smaller radii, the estimated uncertainty is smaller (higher)
    at low (high) $\q2$ than for $R=1$. For $R=1.2$, this uncertainty
    increases up to $\pm 10\%$ for $\q2\approx 500$~\g2;
  \item the uncertainty on the NLO QCD calculations due to those on the 
    proton PDFs was estimated by repeating the calculations using 22
    additional sets from the ZEUS-S analysis, which takes into
    account the statistical and correlated systematic experimental
    uncertainties of each data set used in the determination of the
    proton PDFs. The resulting uncertainty in the cross sections was
    below $\pm 3\%$, except in the high-$\etjb$ region where it reached
    $\pm 4.4\%$;
  \item the uncertainty on the NLO QCD calculations due to that on
    $\asz$ was estimated by repeating the calculations using two
    additional sets of proton PDFs, for which different values of
    $\asz$ were assumed in the fits. The difference between
    the calculations using these various sets was scaled by a factor
    such as to reflect the uncertainty on the current world average of
    $\as$~\cite{jp:g26:r27}. The resulting uncertainty in the
    cross sections was below $\pm 2\%$;
  \item the uncertainty from the modelling of the QCD cascade was
    assumed to be half the difference between the hadronisation
    corrections obtained using the {\sc Ariadne} and {\sc Lepto-MEPS}
    models. The resulting uncertainty on the cross sections was less
    than $1.4\%$ for $R=1$ and increased up to $\sim 4\%$ for $R=0.5$;
  \item the uncertainty of the calculations in the value of $\mu_F$ was
    estimated by repeating the calculations with $\mu_F=Q/2$ and $2Q$.
    The variation of the calculations was negligible.
\end{itemize}
 
The total theoretical uncertainty was obtained by adding in quadrature
the individual uncertainties listed above.

It is concluded that NLO QCD provides predictions with comparable 
precision in the range $R=0.5-1$. For larger values of $R$,
e.g. $R=1.2$, it was estimated that the uncertainty on the NLO QCD
calculations due to terms beyond NLO increases up to about $10\%$ for 
high $\q2$ values. On the other hand, the hadronisation correction
estimated for the cross sections with smaller radii, e.g. $R=0.3$,
increases up to about $40\%$. Therefore, only measurements for the
range $R=0.5-1$ are presented in Section~\ref{results}.

\section{Acceptance corrections}
\label{corfac}
The $\etjb$ and $\q2$ distributions in the data were corrected for
detector effects using bin-by-bin correction factors determined with
the MC samples. These correction factors took into account the
efficiency of the trigger, the selection criteria and the purity and
efficiency of the jet reconstruction. For this approach to be valid, the
uncorrected distributions of the data must be well described by the MC
simulations at the detector level. This condition was satisfied by
both the {\sc Ariadne} and {\sc Lepto}-MEPS MC samples. The average
between the acceptance-correction values obtained with {\sc Ariadne}
and {\sc Lepto}-MEPS was used to correct the data to the hadron
level. The deviations in the results obtained by using either {\sc
  Ariadne} or {\sc Lepto}-MEPS to correct the data from their average
were taken to represent systematic uncertainties of the effect of the
QCD-cascade model in the corrections (see Section~\ref{expunc}). The
acceptance-correction factors differed from unity by typically less than 
$10\%$. 

\subsection{Experimental uncertainties}
\label{expunc}
The following sources of systematic uncertainty were considered for
the measured cross sections:
\begin{itemize}
  \item the uncertainty in the absolute energy scale of the jets was
    estimated to be $\pm 1\%$ for $\etlab>10$~GeV and $\pm 3\%$ for
    lower $\etlab$
    values~\cite{pl:b531:9,epj:c23:615,proc:calor:2002:767}. The
    resulting uncertainty was about $\pm 5\%$;
  \item the differences in the results obtained by using either
    {\sc Ariadne} or {\sc Lepto}-MEPS to correct the data for detector
    effects were taken to represent systematic uncertainties. The
    resulting uncertainty was typically below $\pm 3\%$;
  \item the uncertainty due to the selection cuts was estimated by
    varying the values of the cuts within the resolution of each
    variable; the effect on the cross sections was typically below
    $\pm 3\%$;
  \item the uncertainty on the reconstruction of the boost to the
    Breit frame was estimated by using the direction of the track
    associated to the scattered electron instead of that derived from
    its impact position in the CAL. The effect was typically below
    $\pm 1\%$;
  \item the uncertainty in the absolute energy scale of the electron
    candidate was estimated to be $\pm 1\%$~\cite{epj:c21:443}. The
    resulting uncertainty was below $\pm 1\%$;
  \item the uncertainty in the cross sections due to that in the
    simulation of the trigger was negligible.
\end{itemize}

The systematic uncertainties not associated with the absolute energy
scale of the jets were added in quadrature to the statistical
uncertainties and are shown in the figures as error bars. The
uncertainty due to the absolute energy scale of the jets is shown
separately as a shaded band in each figure, due to the large
bin-to-bin correlation. In addition, there was an overall
normalisation uncertainty of $2.2\%$ from the luminosity
determination, which is not included in the figures.

\section{Results}
\label{results}

\subsection{Inclusive-jet differential cross sections for different
  jet radii}
The inclusive-jet differential cross sections were measured in the
kinematic region $\q2>125$~\g2\ and $|\cgh|<0.65$. These cross
sections include every jet of hadrons in the event with $\etjb>8$~GeV
and $-2<\etajb<1.5$ and were corrected for detector and QED radiative 
effects and the running of $\alpha_{\rm em}$.

\begin{figure}[p]
\vfill
\setlength{\unitlength}{1.0cm}
\begin{picture} (18.0,10.0)
\put (-2.0,0.0){\epsfig{figure=\figdir 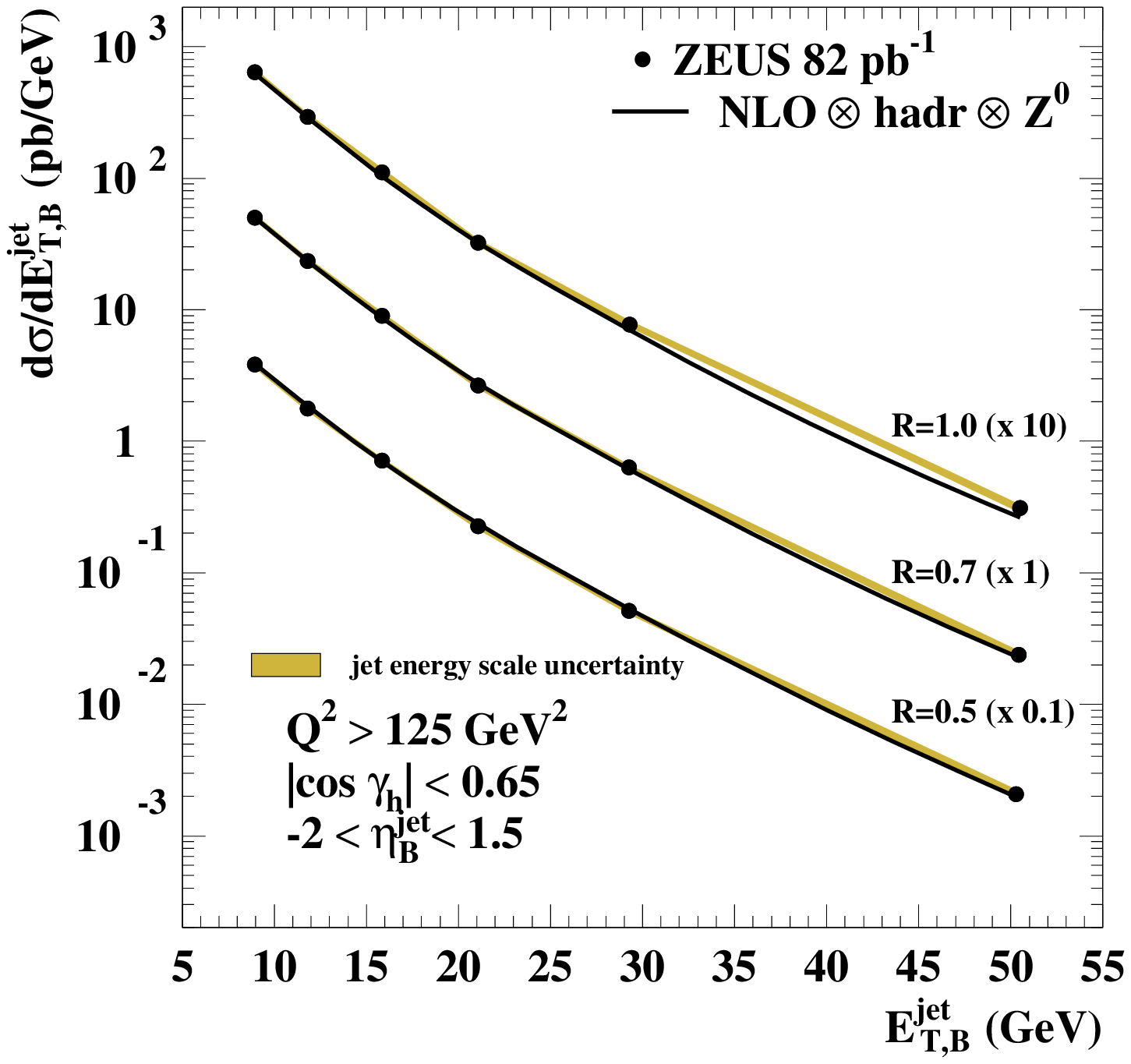,width=12cm}}
\put (7.0,0.0){\epsfig{figure=\figdir 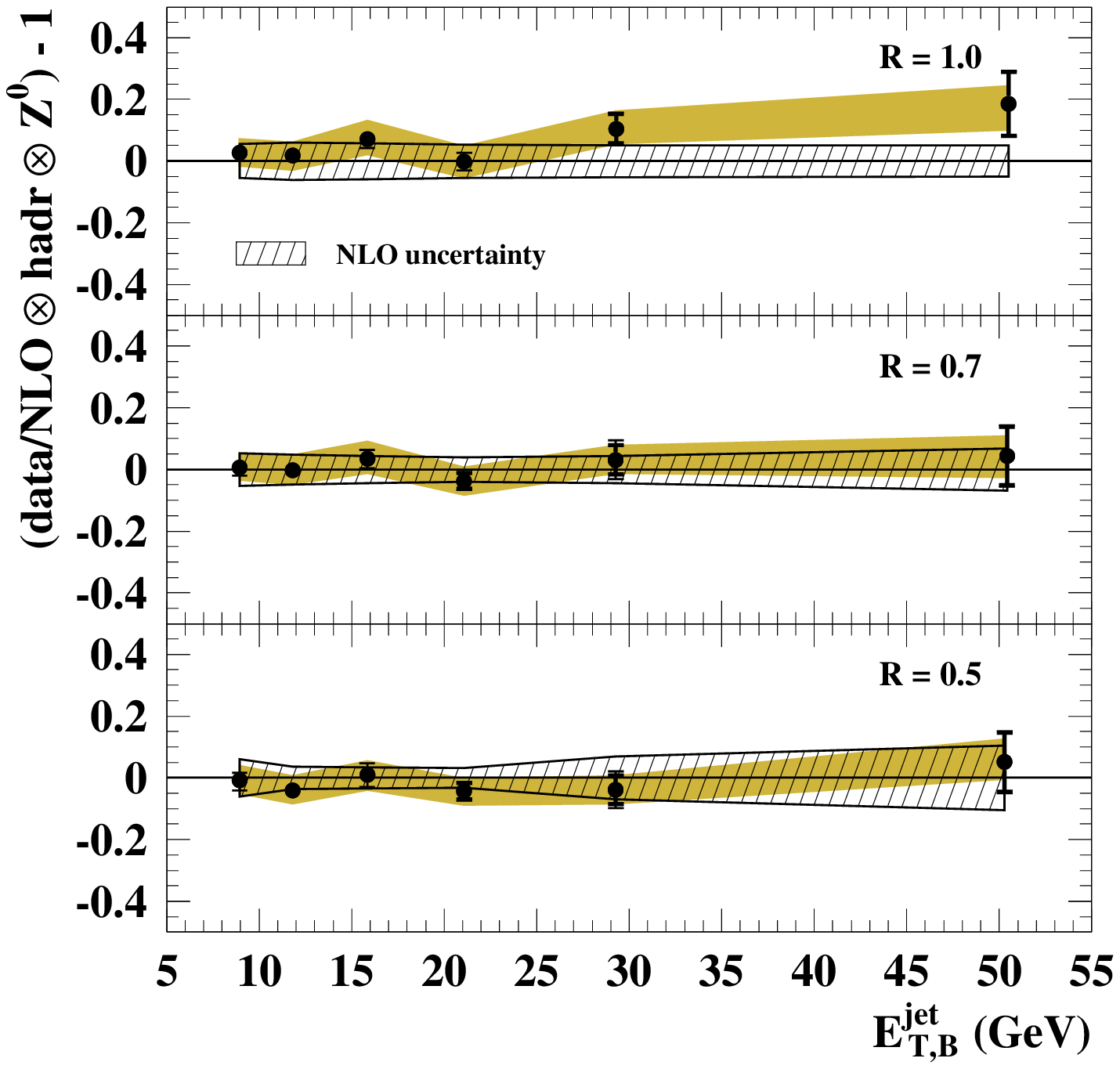,width=12cm}}
\put (1.0,-2.0){\epsfig{figure=\figdir 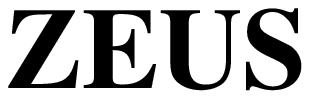,width=15cm}}
\put (6.9,9.2){\bf\small (a)}
\put (15.9,9.2){\bf\small (b)}
\end{picture}
\caption
{\it 
(a) The measured differential cross-section $\setjb$ for inclusive-jet 
production with $-2<\etajb<1.5$ (dots) for different jet radii, in the
kinematic range given by $\q2>125$~\gev$^2$ and $|\cgh|<0.65$.
The NLO QCD calculations with 
$\mu_R=\etjb$ (solid lines), corrected to include hadronisation and $\z0$
effects and using the ZEUS-S parameterisations of the proton PDFs,
are also shown. Each cross section has been multiplied by the scale
factor indicated in brackets to aid visibility. (b) The fractional
differences between the measured $\setjb$ and the NLO QCD calculations
(dots); the hatched bands display the total theoretical
uncertainty.
 The inner
error bars represent the statistical uncertainty. The outer error bars
show the statistical and systematic uncertainties, not associated with 
the uncertainty in the absolute energy scale of the jets, added in 
quadrature. The shaded bands display the uncertainty due to the 
absolute energy scale of the jets. 
}
\label{fig1}
\vfill
\end{figure}

\begin{figure}[p]
\vfill
\setlength{\unitlength}{1.0cm}
\begin{picture} (18.0,10.0)
\put (-2.0,0.0){\epsfig{figure=\figdir 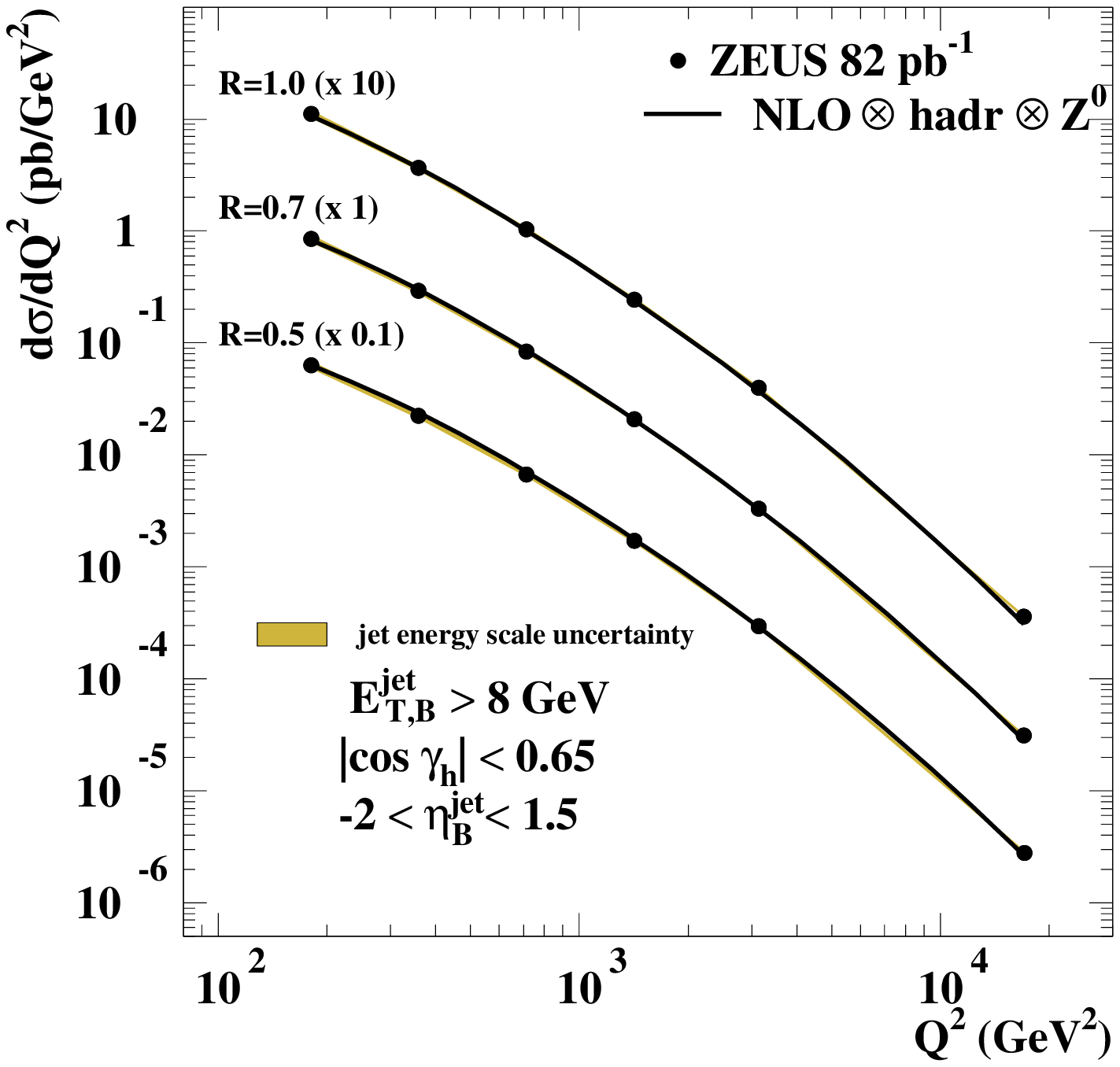,width=12cm}}
\put (7.0,0.0){\epsfig{figure=\figdir 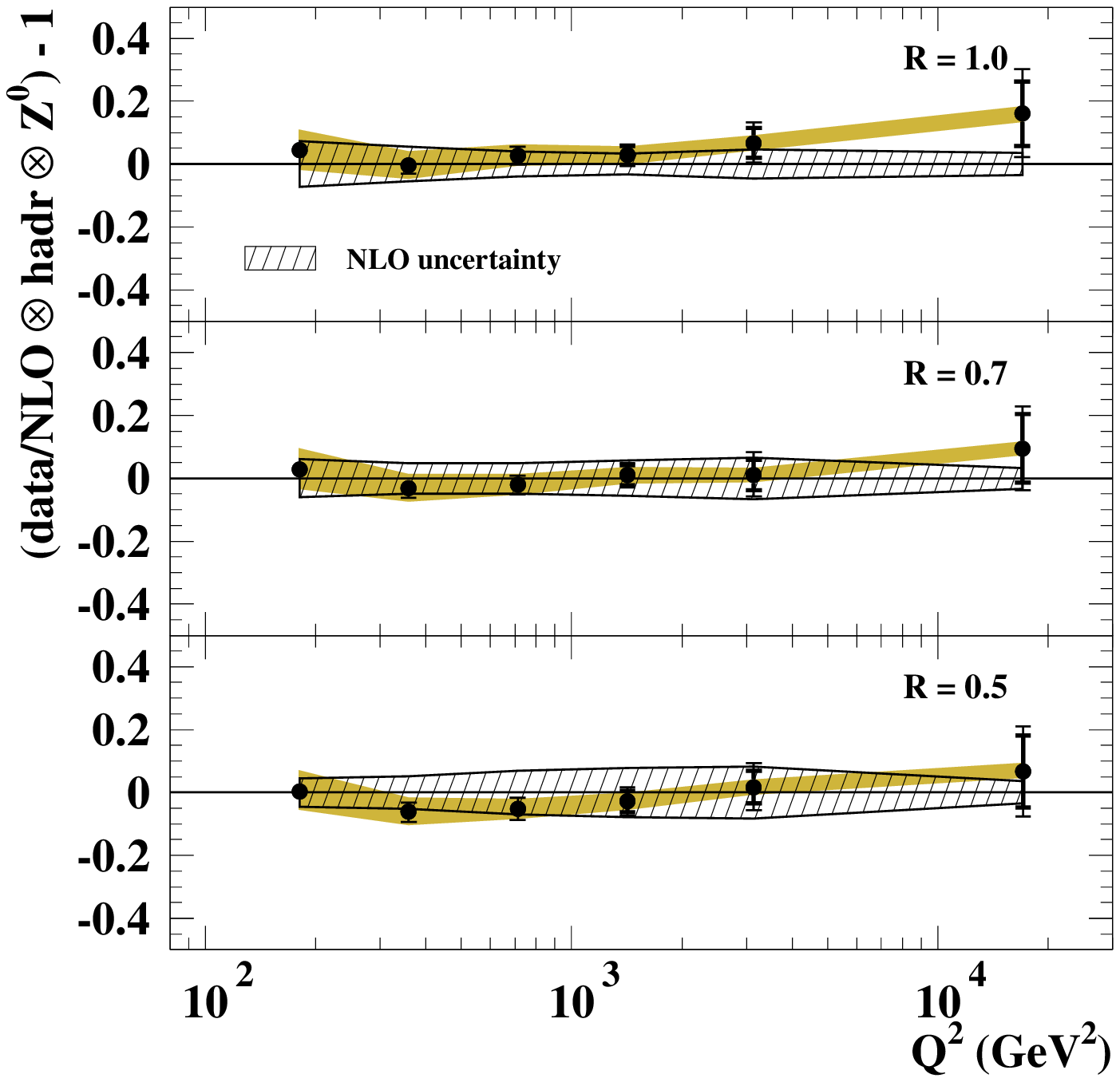,width=12cm}}
\put (1.0,-2.0){\epsfig{figure=\figdir zeus.eps,width=15cm}}
\put (6.9,9.2){\bf\small (a)}
\put (15.9,9.2){\bf\small (b)}
\end{picture}
\caption
{\it 
The measured differential cross-section $\sq2$ for inclusive-jet 
production with $\etjb>8$~GeV and $-2<\etajb<1.5$ (dots) for different
jet radii, in the kinematic range given by $|\cgh|<0.65$. Other 
details as in the caption to Fig.~\ref{fig1}. 
}
\label{fig2}
\vfill
\end{figure}

The measurements of the inclusive-jet differential cross sections as
functions of $\etjb$ and $\q2$ are presented in Figs.~\ref{fig1}a and
\ref{fig2}a for jet radii $R=1$, $0.7$ and $0.5$. In these figures,
each data point is plotted at the abscissa at which the NLO QCD
differential cross section was equal to its bin-averaged value. The
measured $\setjb$ ($\sq2$) exhibits a steep fall-off over three (five)
orders of magnitude for the jet radii considered in the $\etjb$
($\q2$) measured range.

The NLO QCD predictions with $\mu_R=\etjb$ are compared to the
measurements in Figs.~\ref{fig1}a and \ref{fig2}a. The fractional
difference of the measured differential cross sections to the NLO QCD
calculations is shown in Figs.~\ref{fig1}b and \ref{fig2}b. The
calculations reproduce the measured differential cross sections well
for the jet radii considered, with similar precision. To study the
scale dependence, NLO QCD calculations using $\mu_R=Q$ were also
compared to the data (not shown); they also provide a good description
of the data.
 
\subsection{Dependence of the inclusive-jet cross section on the jet
  radius}
\begin{figure}[p]
\vfill
\setlength{\unitlength}{1.0cm}
\begin{picture} (18.0,15.0)
\put (-2.0,0.0){\epsfig{figure=\figdir 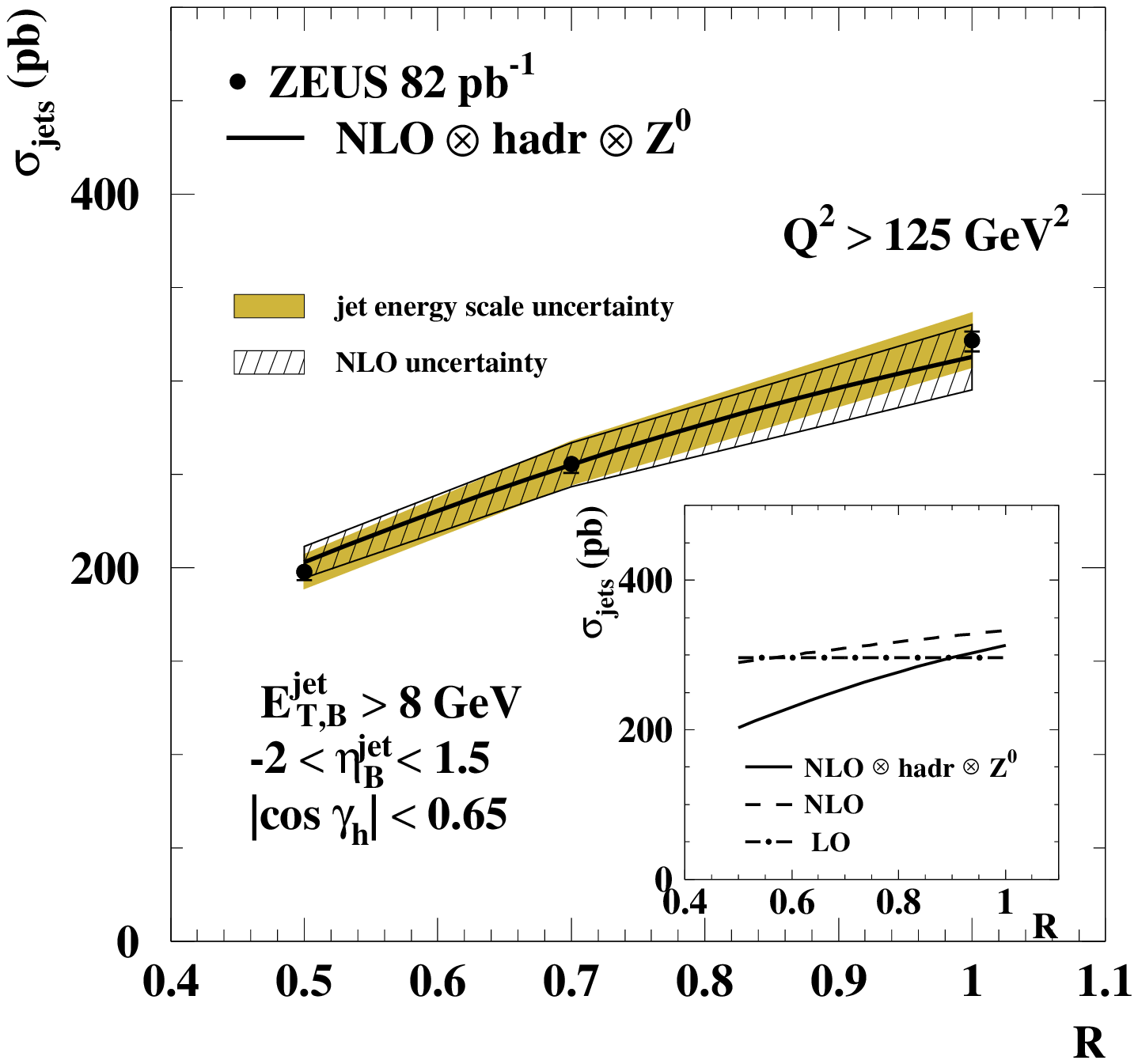,width=12cm}}
\put (7.0,0.0){\epsfig{figure=\figdir 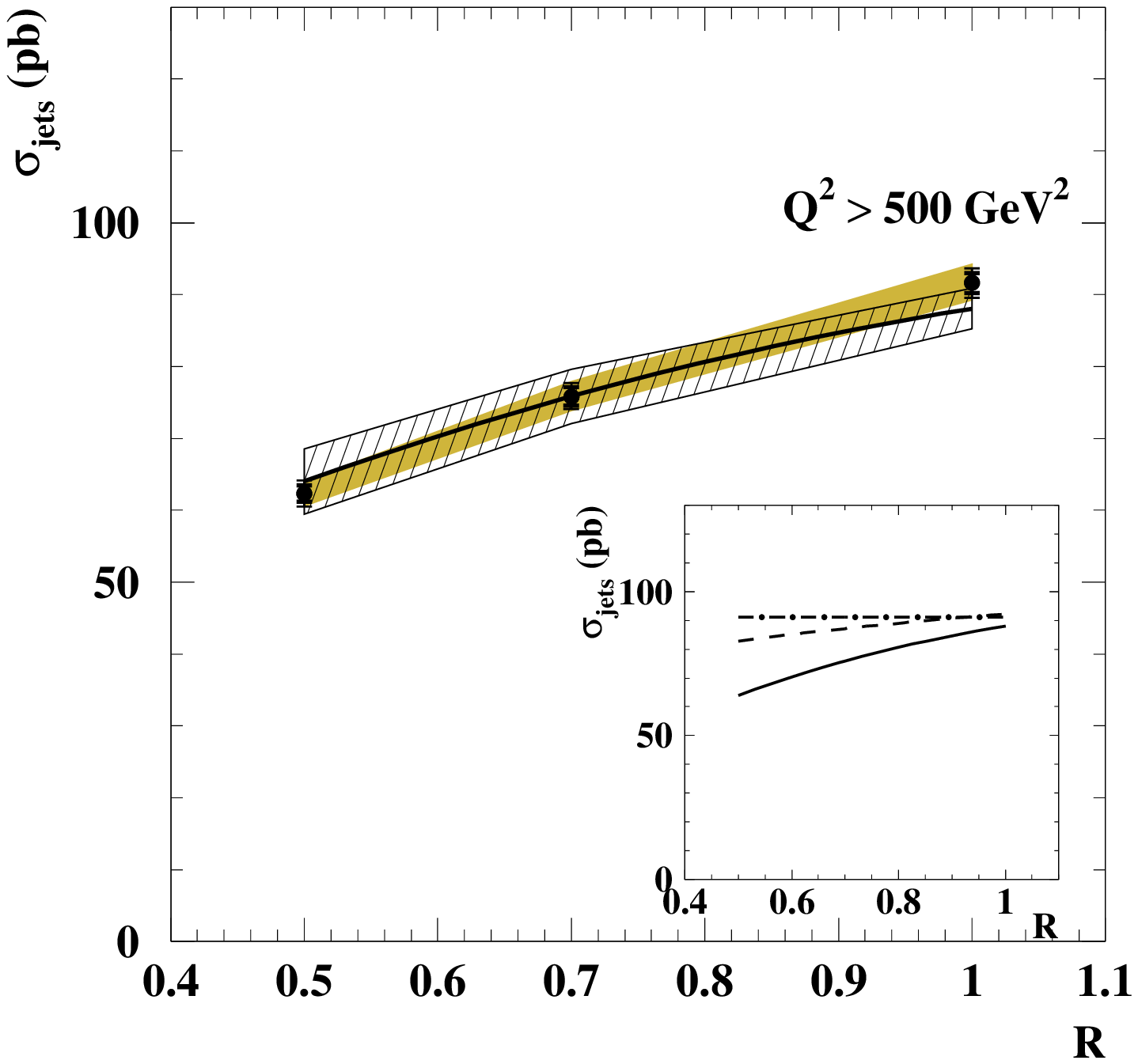,width=12cm}}
\put (1.0,-2.0){\epsfig{figure=\figdir zeus.eps,width=15cm}}
\put (6.7,9.0){\bf\small (a)}
\put (15.7,9.0){\bf\small (b)}
\end{picture}
\caption
{\it 
The measured cross-section $\sigma_{\rm jets}$ as a function of the 
jet radius for inclusive-jet production with $\etjb>8$~GeV and
$-2<\etajb<1.5$ (dots), in the kinematic range given by $|\cgh|<0.65$
and (a) $\q2>125$~\gev$^2$\ and (b) $\q2>500$~\gev$^2$. The insets show the
LO (dot-dashed lines) and NLO (dashed lines) QCD calculations. The NLO
QCD calculations corrected to include hadronisation and $\z0$ effects are
shown as solid lines. Other details as in the caption to
Fig.~\ref{fig1}.
}
\label{fig3}
\vfill
\end{figure}

Measurements of the inclusive-jet cross section have been performed
for $\etjb>8$~GeV and $-2<\etajb<1.5$ in the kinematic range given by
$|\cgh|<0.65$ integrated above $\q2_{\rm min}=125$ and $500$~\g2\ for
different jet radii. The measured cross section, $\sigma_{\rm jets}$, as
a function of $R$ is presented in Figs.~\ref{fig3}a and
\ref{fig3}b. The measured cross sections increase linearly
with $R$ in the range between $0.5$ and $1$. The increase of 
$\sigma_{\rm jets}$ as $R$ increases can be understood as
the result of more transverse energy being gathered in a jet so that
a larger number of jets has $\etjb$ exceeding the threshold of $8$~GeV.

The predictions of LO and NLO QCD for $\sigma_{\rm jets}$
at the parton level, with no corrections for hadronisation or $\z0$
exchange, are shown in the inset of Fig.~\ref{fig3}. The LO
predictions do not depend on $R$ since there is only one parton per
jet. The NLO calculations give the lowest-order contribution to the
$R$ dependence of the inclusive-jet cross section. 
The NLO QCD calculations, corrected to include hadronisation and $\z0$
effects, are also shown in the inset of Fig.~\ref{fig3}. It is
observed that, in addition to the effects due to parton radiation, the
hadronisation corrections further modify the shape of the prediction.

The NLO QCD calculations are compared to the data in
Fig.~\ref{fig3}. They give a good description of the $R$ dependence of
the data within the jet-radius range considered. The uncertainty of
the NLO calculation is also shown. The total theoretical
uncertainty of the cross section integrated above 
$\q2_{\rm min}=125$~\g2\ ($500$~\g2) changes from $5.6\%$ ($3.2\%$)
for $R=1$ to $4.2\%$ ($7.1\%$) for $R=0.5$. This variation is due to a
change of the uncertainty arising from higher-order QCD corrections
and hadronisation. The uncertainties arising from $\asz$ and the
proton PDFs do not change significantly with $R$.

\subsection{Determination of {\boldmath $\asz$}}
The measured differential cross sections presented in Section~7.1
were used to determine a value of $\asz$ using the method 
presented previously~\cite{pl:b547:164}. The NLO QCD calculations were
performed using the program {\sc Disent} with five sets of
ZEUS-S proton PDFs which were determined from global fits assuming
different values of $\asz$, namely 
$\asz=0.115,\ 0.117,\ 0.119,\ 0.121$ and $0.123$. The value of $\asz$
used in each calculation was that associated
with the corresponding set of PDFs. The $\asz$ dependence of the
predicted cross sections in each bin $i$ of $A$ ($A=\q2,\etjb$) was 
parameterised according to 
$$\left [ d\sigma/dA(\asz) \right ]_i=C_1^i\asz+C_2^i\as^2(\mz),$$
where $C_1^i$ and $C_2^i$ were determined from a $\chi^2$ fit to
the NLO QCD calculations. The value of $\asz$ was determined by a
$\chi^2$ fit to the measured $d\sigma/dA$ values for several regions
of the variable $A$. The values of $\asz$ obtained from the various
differential cross sections and jet radii were found to be
consistent. The result obtained using the measured $\sq2$ for
$\q2>500$~\g2\ with $R=1$ yields the smallest uncertainty and, therefore,
represents the most precise determination from this analysis.
In the region $\q2>500$~\g2, the experimental systematic uncertainties
on the cross sections are smaller than at lower $\q2$ and the
theoretical uncertainties due to the proton PDFs and to terms beyond
NLO are minimised.

The uncertainties on the extracted values of $\asz$ due to the
experimental systematic uncertainties were evaluated by repeating the
analysis for each systematic check presented in Section~\ref{expunc}.
The overall normalisation uncertainty from the luminosity determination
was also considered.
The largest contribution to the experimental uncertainty comes from the
jet energy scale and amounts to $\pm 2\%$ on $\asz$. The theoretical
uncertainties were evaluated as described in Section~\ref{thun}. The
largest contribution was the theoretical uncertainty arising from
terms beyond NLO, which was estimated by using the method proposed by
Jones \etal~\cite{jhep:0312:007}, and amounted to $\pm 1.5\%$. The
uncertainty due to the proton PDFs was $\pm 0.7\%$. The uncertainty
arising from the hadronisation effects amounted to $\pm 0.8\%$.

As a cross-check, $\asz$ was determined by using NLO QCD
calculations based on the MRST2001~\cite{epj:c28:455} and
CTEQ6~\cite{jhep:0207:012,*jhep:0310:046} sets of proton PDFs. The values
obtained are consistent within $1\%$ with those based on ZEUS-S. The
uncertainty arising from the proton PDFs was estimated to be $\pm 0.7\%$ 
($\pm 1.6\%$) using the results of the MRST2001 (CTEQ6) analysis.

The value of $\asz$ obtained from the measured $\sq2$ for
$\q2>500$~\g2\ with $R=1$ is
\begin{center}
$\asmz{0.1207}{0.0014}{0.0033}{0.0035}{0.0023}{0.0022}$.
\end{center}

This value of $\asz$ is consistent with the current world 
average of $0.1189\pm 0.0010$~\cite{jp:g26:r27} as well as with the
HERA average of $0.1186\pm 0.0051$~\cite{proc:dis:2005:689}. It has a
precision comparable to the values obtained from $\ele$
interactions~\cite{jp:g26:r27}.

\subsection{Energy-scale dependence of $\as$}
\begin{figure}[p]
\vfill
\setlength{\unitlength}{1.0cm}
\begin{picture} (18.0,18.0)
\put (-1.0,0.0){\epsfig{figure=\figdir 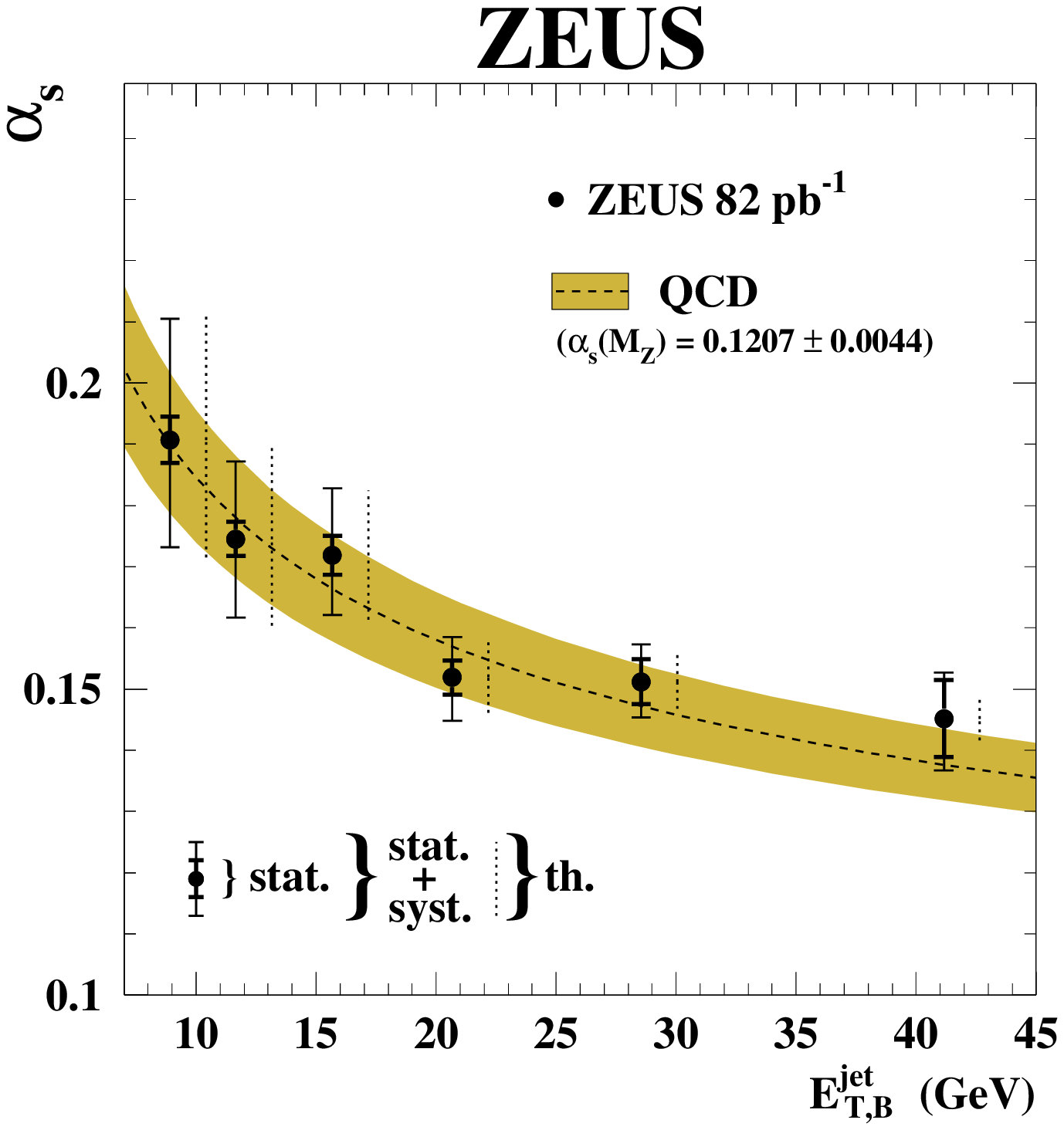,width=18cm}}
\end{picture}
\vspace{-1.5cm}
\caption
{\it 
The $\as$ values determined from the
measured $\setjb$ with $R=1$ as a function of $\etjb$ (dots).
The dashed line indicates the renormalisation group prediction
at two loops obtained from the $\as(\mz)$ value determined in this
analysis and  the shaded area represents its uncertainty. The inner 
error bars represent the statistical uncertainties of the data. The 
outer error bars show the statistical and systematic
uncertainties added in quadrature. The dotted vertical bars, shifted
to aid visibility, represent the theoretical uncertainties.
}
\label{fig4}
\vfill
\end{figure}

The QCD prediction for the energy-scale dependence of the strong
coupling constant was tested by determining $\as$ from the measured
$\setjb$ with $R=1$ at different $\etjb$ values. The method employed
was the same as described above, but parameterising the $\as$
dependence of $\setjb$ in terms of $\as(\langle\etjb\rangle)$ instead
of $\asz$, where $\langle\etjb\rangle$ is the average $\etjb$ of the
data in each bin. The extracted values of $\as$ are shown in
Fig.~\ref{fig4}. The results are in good agreement with the predicted
running of the strong coupling
constant~\cite{prl:30:1343,*prl:30:1346,*pr:d8:3633,*prep:14:129}
calculated at two loops~\cite{prl:33:244,*np:b75:531,*tmf:41:26} over
a large range in $\etjb$.

\section{Summary}
Measurements of differential cross sections for inclusive-jet
production in neutral current deep inelastic $ep$ scattering at a
centre-of-mass energy of 318~GeV have been presented. The cross
sections refer to jets of hadrons identified in the Breit frame with
the $\kt$ cluster algorithm in the longitudinally invariant inclusive
mode. The cross sections are given in the kinematic region of
$\q2>125$~\g2\ and $|\cgh|<0.65$.

The dependence of the inclusive-jet cross sections on the jet-radius
$R$ has been studied. It has been determined that NLO QCD provides
predictions with comparable precision in the range $R=0.5-1$. 
Measurements of inclusive-jet differential cross sections for this
jet-radius range have been presented. The NLO QCD calculations provide
a good description of the measured inclusive-jet differential cross
sections $\setjb$ and $\sq2$ for $\RRR$. It is observed that the
measured inclusive-jet cross section integrated above 
$\q2_{\rm min}=125$ and $500$~\g2\ increases linearly with $R$ in the
jet-radius range studied.

The measured inclusive-jet differential cross sections have been used
to extract a value of $\asz$. A QCD fit of the cross-section
$\sq2$ with $R=1$ for $\q2>500$~\g2\  yields the determination with
smallest uncertainty, 
\begin{center}
$\asmz{0.1207}{0.0014}{0.0033}{0.0035}{0.0023}{0.0022}$.
\end{center}

This value is in good agreement with the world and HERA averages.
The extracted values of $\as$ at different $\etjb$ are in good
agreement with the predicted running of the strong coupling constant
over a large range in $\etjb$.

\vspace{0.5cm}
\noindent {\Large\bf Acknowledgements}
\vspace{0.3cm}

We thank the DESY Directorate for their strong support and
encouragement. The remarkable achievements of the HERA machine group
were essential for the successful completion of this work and are
greatly appreciated. We are grateful for the support of the DESY
computing and network services. The design, construction and
installation of the ZEUS detector have been made possible owing to the
ingenuity and effort of many people who are not listed as authors.

\vfill\eject

\providecommand{\etal}{et al.\xspace}
\providecommand{\coll}{Collaboration}
\catcode`\@=11
\def\@bibitem#1{%
\ifmc@bstsupport
  \mc@iftail{#1}%
    {;\newline\ignorespaces}%
    {\ifmc@first\else.\fi\orig@bibitem{#1}}
  \mc@firstfalse
\else
  \mc@iftail{#1}%
    {\ignorespaces}%
    {\orig@bibitem{#1}}%
\fi}%
\catcode`\@=12

\clearpage
\newpage
\begin{table}
\begin{center}
    \begin{tabular}{||c|cccc||c||c||}
\hline
  $\etjb$ bin
& $d\sigma/d\etjb$
&
&
&
&
& \\
  (GeV)
& (pb/GeV)
& $\delta_{\rm stat}$
& $\delta_{\rm syst}$
& $\delta_{\rm ES}$
& $C_{\rm QED}$
& $C_{\rm had}$\\
\hline
\multicolumn{7}{||c||}{$R=1$} \\
\hline
 8-10
&63.98
&0.68
&{\small ${}_{-1.39}^{+1.20}$}
&{\small ${}_{-2.56}^{+2.84}$}
&0.95
&0.91
\\
 10-14
&29.29
&0.34
&{\small ${}_{-0.52}^{+0.37}$}
&{\small ${}_{-1.31}^{+1.25}$}
&0.96
&0.95
\\
 14-18
&11.07
&0.20
&{\small ${}_{-0.20}^{+0.15}$}
&{\small ${}_{-0.50}^{+0.61}$}
&0.96
&0.96
\\
 18-25
&3.234
&0.080
&{\small ${}_{-0.045}^{+0.036}$}
&{\small ${}_{-0.167}^{+0.156}$}
&0.94
&0.97
\\
 25-35
&0.773
&0.033
&{\small ${}_{-0.015}^{+0.016}$}
&{\small ${}_{-0.033}^{+0.040}$}
&0.95
&0.95
\\
 35-100
&0.0312
&0.0027
&{\small ${}_{-0.0006}^{+0.0005}$}
&{\small ${}_{-0.0022}^{+0.0015}$}
&1.06
&0.95
\\
\hline
\multicolumn{7}{||c||}{$R=0.7$} \\
\hline
 8-10
&50.09
&0.60
&{\small ${}_{-1.19}^{+0.93}$}
&{\small ${}_{-1.94}^{+2.16}$}
&0.95
&0.77
\\
 10-14
&23.38
&0.30
&{\small ${}_{-0.38}^{+0.30}$}
&{\small ${}_{-1.06}^{+1.11}$}
&0.96
&0.83
\\
 14-18
&8.97
&0.18
&{\small ${}_{-0.19}^{+0.17}$}
&{\small ${}_{-0.41}^{+0.47}$}
&0.96
&0.88
\\
 18-25
&2.659
&0.071
&{\small ${}_{-0.038}^{+0.031}$}
&{\small ${}_{-0.126}^{+0.120}$}
&0.95
&0.91
\\
 25-35
&0.631
&0.029
&{\small ${}_{-0.025}^{+0.024}$}
&{\small ${}_{-0.026}^{+0.028}$}
&0.96
&0.92
\\
 35-100
&0.0237
&0.0022
&{\small ${}_{-0.0007}^{+0.0006}$}
&{\small ${}_{-0.0016}^{+0.0015}$}
&1.03
&0.93
\\
\hline
\multicolumn{7}{||c||}{$R=0.5$} \\
\hline
 8-10
&38.25
&0.52
&{\small ${}_{-1.11}^{+0.85}$}
&{\small ${}_{-1.60}^{+1.81}$}
&0.95
&0.64
\\
 10-14
&17.78
&0.26
&{\small ${}_{-0.30}^{+0.26}$}
&{\small ${}_{-0.77}^{+0.85}$}
&0.96
&0.70
\\
 14-18
&7.09
&0.15
&{\small ${}_{-0.23}^{+0.22}$}
&{\small ${}_{-0.33}^{+0.32}$}
&0.95
&0.77
\\
 18-25
&2.257
&0.063
&{\small ${}_{-0.032}^{+0.027}$}
&{\small ${}_{-0.103}^{+0.096}$}
&0.96
&0.83
\\
 25-35
&0.514
&0.025
&{\small ${}_{-0.020}^{+0.020}$}
&{\small ${}_{-0.024}^{+0.024}$}
&0.97
&0.86
\\
 35-100
&0.0208
&0.0019
&{\small ${}_{-0.0006}^{+0.0006}$}
&{\small ${}_{-0.0011}^{+0.0014}$}
&1.04
&0.90
\\
\hline
    \end{tabular}
 \caption{
   Inclusive jet cross-sections $d\sigma/d\etjb$ for jets of hadrons
   in the Breit frame selected with the longitudinally invariant $\kt$
   cluster algorithm for different values of $R$ (Fig. 1). The
   statistical, uncorrelated systematic and jet-energy-scale ({\rm ES})
   uncertainties are shown separately. The multiplicative corrections
   applied to the data to correct for QED radiative effects, 
   $C_{\rm QED}$, and the corrections for hadronisation effects to be
   applied to the parton-level NLO QCD calculations, $C_{\rm had}$,
   are shown in the last two columns.}
 \label{tabone}
\end{center}
\end{table}

\clearpage
\newpage
\begin{table}
\begin{center}
    \begin{tabular}{||c|cccc||c||c||}
\hline
  $\q2$ bin
& $d\sigma/d\q2$
&
&
&
&
& \\
  (\g2)
& (pb/\g2)
& $\delta_{\rm stat}$
& $\delta_{\rm syst}$
& $\delta_{\rm ES}$
& $C_{\rm QED}$
& $C_{\rm had}$\\
\hline
\multicolumn{7}{||c||}{$R=1$} \\
\hline
 125-250
&1.106
&0.012
&{\small ${}_{-0.020}^{+0.013}$}
&{\small ${}_{-0.062}^{+0.066}$}
&0.97
&0.92
\\
 250-500
&0.3671
&0.0053
&{\small ${}_{-0.0078}^{+0.0048}$}
&{\small ${}_{-0.0149}^{+0.0153}$}
&0.95
&0.94
\\
 500-1000
&0.1037
&0.0020
&{\small ${}_{-0.0021}^{+0.0020}$}
&{\small ${}_{-0.0029}^{+0.0033}$}
&0.95
&0.95
\\
 1000-2000
&0.02439
&0.00072
&{\small ${}_{-0.00033}^{+0.00039}$}
&{\small ${}_{-0.00058}^{+0.00059}$}
&0.94
&0.96
\\
 2000-5000
&0.00396
&0.00017
&{\small ${}_{-0.00015}^{+0.00017}$}
&{\small ${}_{-0.00008}^{+0.00008}$}
&0.94
&0.95
\\
 5000-100000
&0.000036
&0.000003
&{\small ${}_{-0.000003}^{+0.000003}$}
&{\small ${}_{-0.000001}^{+0.000001}$}
&0.98
&0.96
\\
\hline
\multicolumn{7}{||c||}{$R=0.7$} \\
\hline
 125-250
&0.855
&0.010
&{\small ${}_{-0.012}^{+0.007}$}
&{\small ${}_{-0.048}^{+0.054}$}
&0.97
&0.79
\\
 250-500
&0.2913
&0.0046
&{\small ${}_{-0.0079}^{+0.0053}$}
&{\small ${}_{-0.0119}^{+0.0124}$}
&0.95
&0.83
\\
 500-1000
&0.0840
&0.0018
&{\small ${}_{-0.0017}^{+0.0017}$}
&{\small ${}_{-0.0024}^{+0.0026}$}
&0.95
&0.86
\\
 1000-2000
&0.02079
&0.00066
&{\small ${}_{-0.00043}^{+0.00041}$}
&{\small ${}_{-0.00049}^{+0.00046}$}
&0.94
&0.88
\\
 2000-5000
&0.00332
&0.00016
&{\small ${}_{-0.00016}^{+0.00018}$}
&{\small ${}_{-0.00006}^{+0.00007}$}
&0.93
&0.88
\\
 5000-100000
&0.000031
&0.000003
&{\small ${}_{-0.000002}^{+0.000002}$}
&{\small ${}_{-0.000001}^{+0.000001}$}
&0.97
&0.90
\\
\hline
\multicolumn{7}{||c||}{$R=0.5$} \\
\hline
 125-250
&0.6344
&0.0088
&{\small ${}_{-0.0092}^{+0.0058}$}
&{\small ${}_{-0.0357}^{+0.0406}$}
&0.97
&0.64
\\
 250-500
&0.2246
&0.0040
&{\small ${}_{-0.0069}^{+0.0053}$}
&{\small ${}_{-0.0097}^{+0.0097}$}
&0.95
&0.70
\\
 500-1000
&0.0672
&0.0016
&{\small ${}_{-0.0019}^{+0.0020}$}
&{\small ${}_{-0.0020}^{+0.0021}$}
&0.94
&0.75
\\
 1000-2000
&0.01709
&0.00060
&{\small ${}_{-0.00051}^{+0.00049}$}
&{\small ${}_{-0.00042}^{+0.00043}$}
&0.94
&0.79
\\
 2000-5000
&0.00296
&0.00015
&{\small ${}_{-0.00015}^{+0.00016}$}
&{\small ${}_{-0.00006}^{+0.00006}$}
&0.95
&0.81
\\
 5000-100000
&0.000028
&0.000003
&{\small ${}_{-0.000002}^{+0.000002}$}
&{\small ${}_{-0.000000}^{+0.000001}$}
&0.98
&0.83
\\
\hline
    \end{tabular}
 \caption{
   Inclusive jet cross-sections $d\sigma/d\q2$ for jets of hadrons
   in the Breit frame selected with the longitudinally invariant $\kt$
   cluster algorithm for different values of $R$ (Fig. 2). Other
   details as in the caption to Table~\ref{tabone}.}
 \label{tabtwo}
\end{center}
\end{table}

\clearpage
\newpage
\begin{table}
\begin{center}
    \begin{tabular}{||c|cccc||c||c||}
\hline
  $R$
& $\sigma_{\rm jets}$
&
&
&
&
& \\
  
& (pb)
& $\delta_{\rm stat}$
& $\delta_{\rm syst}$
& $\delta_{\rm ES}$
& $C_{\rm QED}$
& $C_{\rm had}$\\
\hline
\multicolumn{7}{||c||}{$\q2>125$~\g2} \\
\hline
 0.5
&197.8
&1.9
&{\small ${}_{-4.1}^{+3.3}$}
&{\small ${}_{-8.6}^{+9.3}$}
&0.96
&0.70
\\
 0.7
&255.6
&2.1
&{\small ${}_{-4.4}^{+3.3}$}
&{\small ${}_{-11.1}^{+11.9}$}
&0.96
&0.82
\\
 1.0
&321.5
&2.4
&{\small ${}_{-5.4}^{+4.2}$}
&{\small ${}_{-14.1}^{+14.8}$}
&0.96
&0.94
\\
\hline
\multicolumn{7}{||c||}{$\q2>500$~\g2} \\
\hline
 0.5
&62.3
&1.1
&{\small ${}_{-1.4}^{+1.5}$}
&{\small ${}_{-1.7}^{+1.7}$}
&0.95
&0.77
\\
 0.7
&75.8
&1.3
&{\small ${}_{-1.3}^{+1.3}$}
&{\small ${}_{-2.0}^{+2.1}$}
&0.95
&0.87
\\
 1.0
&91.6
&1.4
&{\small ${}_{-1.5}^{+1.6}$}
&{\small ${}_{-2.4}^{+2.6}$}
&0.95
&0.95
\\
\hline
    \end{tabular}
 \caption{
   Inclusive jet cross-sections $\sigma_{\rm jets}$ for jets of hadrons
   in the Breit frame selected with the longitudinally invariant $\kt$
   cluster algorithm for $\q2>125$ and $500$~\gev$^2$ (Fig. 3). Other
   details as in the caption to Table~\ref{tabone}.}
 \label{tabthree}
\end{center}
\end{table}

\begin{table}
\begin{center}
    \begin{tabular}{||c|cccc||}
\hline
  $\langle\etjb\rangle$
& $\as$
&
&
& \\
  (GeV)
& 
& $\delta_{\rm stat}$
& $\delta_{\rm syst}$
& $\delta_{\rm theor}$\\
\hline
 8.9
&0.1907
&{\small ${}_{-0.0038}^{+0.0038}$}
&{\small ${}_{-0.0171}^{+0.0194}$}
&{\small ${}_{-0.0192}^{+0.0208}$}
\\
 11.7
&0.1746
&{\small ${}_{-0.0028}^{+0.0028}$}
&{\small ${}_{-0.0126}^{+0.0123}$}
&{\small ${}_{-0.0142}^{+0.0148}$}
\\
 15.7
&0.1719
&{\small ${}_{-0.0031}^{+0.0032}$}
&{\small ${}_{-0.0092}^{+0.0105}$}
&{\small ${}_{-0.0105}^{+0.0107}$}
\\
 20.7
&0.1519
&{\small ${}_{-0.0028}^{+0.0028}$}
&{\small ${}_{-0.0065}^{+0.0061}$}
&{\small ${}_{-0.0057}^{+0.0057}$}
\\
 28.6
&0.1512
&{\small ${}_{-0.0037}^{+0.0037}$}
&{\small ${}_{-0.0045}^{+0.0050}$}
&{\small ${}_{-0.0044}^{+0.0043}$}
\\
 41.2
&0.1452
&{\small ${}_{-0.0063}^{+0.0064}$}
&{\small ${}_{-0.0056}^{+0.0041}$}
&{\small ${}_{-0.0036}^{+0.0036}$}
\\
\hline
    \end{tabular}
 \caption{
   The $\as$ values determined from a QCD fit of the measured $\setjb$
   with $R=1$ as a function of $\etjb$ (Fig. 4). The statistical,
   systematic and theoretical uncertainties are shown
   separately.}
 \label{tabfour}
\end{center}
\end{table}

\newpage
\clearpage

\newpage
\clearpage

\newpage
\clearpage

\newpage
\clearpage


\begin{mcbibliography}{10}

\bibitem{pl:b507:70}
\colab{ZEUS}, J. Breitweg \etal,
\newblock Phys.\ Lett.{} B~507~(2001)~70\relax
\relax
\bibitem{epj:c19:289}
\colab{H1}, C. Adloff \etal,
\newblock Eur.\ Phys.\ J.{} C~19~(2001)~289\relax
\relax
\bibitem{epj:c23:13}
\colab{ZEUS}, S. Chekanov \etal,
\newblock Eur.\ Phys.\ J.{} C~23~(2002)~13\relax
\relax
\bibitem{hep-ex-0608048}
\colab{ZEUS}, S. Chekanov \etal,
\newblock Preprint \mbox{DESY-06-128} (\mbox{hep-ex/0608048}), DESY, 2006\relax
\relax
\bibitem{pl:b547:164}
\colab{ZEUS}, S. Chekanov \etal,
\newblock Phys.\ Lett.{} B~547~(2002)~164\relax
\relax
\bibitem{pl:b551:226}
\colab{ZEUS}, S. Chekanov \etal,
\newblock Phys.\ Lett.{} B~551~(2003)~226\relax
\relax
\bibitem{pl:b515:17}
\colab{H1}, C. Adloff \etal,
\newblock Phys.\ Lett.{} B~515~(2001)~17\relax
\relax
\bibitem{epj:c44:183}
\colab{ZEUS}, S. Chekanov \etal,
\newblock Eur.\ Phys.\ J.{} C~44~(2005)~183\relax
\relax
\bibitem{epj:c11:35}
\colab{ZEUS}, J.~Breitweg \etal,
\newblock Eur.\ Phys.\ J.{} C~11~(1999)~35\relax
\relax
\bibitem{epj:c23:615}
\colab{ZEUS}, S. Chekanov \etal,
\newblock Eur.\ Phys.\ J.{} C~23~(2002)~615\relax
\relax
\bibitem{pl:b531:9}
\colab{ZEUS}, S. Chekanov \etal,
\newblock Phys.\ Lett.{} B~531~(2002)~9\relax
\relax
\bibitem{epj:c25:13}
\colab{H1}, C.~Adloff \etal,
\newblock Eur.\ Phys.\ J.{} C~25~(2002)~13\relax
\relax
\bibitem{pl:b639:21}
\colab{H1}, A. Aktas \etal,
\newblock Phys.\ Lett.{} B~639~(2006)~21\relax
\relax
\bibitem{pl:b560:7}
\colab{ZEUS}, S. Chekanov \etal,
\newblock Phys.\ Lett.{} B~560~(2003)~7\relax
\relax
\bibitem{epj:c29:497}
\colab{H1}, C. Adloff \etal,
\newblock Eur.\ Phys.\ J.{} C~29~(2003)~497\relax
\relax
\bibitem{pl:b443:394}
\colab{ZEUS}, J.~Breitweg \etal,
\newblock Phys.\ Lett.{} B~443~(1998)~394\relax
\relax
\bibitem{np:b545:3}
\colab{H1}, C. Adloff \etal,
\newblock Nucl.\ Phys.{} B~545~(1999)~3\relax
\relax
\bibitem{pl:b558:41}
\colab{ZEUS}, S. Chekanov \etal,
\newblock Phys.\ Lett.{} B~558~(2003)~41\relax
\relax
\bibitem{np:b700:3}
\colab{ZEUS}, S. Chekanov \etal,
\newblock Nucl.\ Phys.{} B~700~(2004)~3\relax
\relax
\bibitem{epj:c31:149}
\colab{ZEUS}, S.~Chekanov \etal,
\newblock Eur.\ Phys.\ J.{} C~31~(2003)~149\relax
\relax
\bibitem{np:b406:187}
S. Catani \etal,
\newblock Nucl.\ Phys.{} B~406~(1993)~187\relax
\relax
\bibitem{pr:d48:3160}
S.D. Ellis and D.E. Soper,
\newblock Phys.\ Rev.{} D~48~(1993)~3160\relax
\relax
\bibitem{zfp:c62:127}
M.H. Seymour,
\newblock Z.\ Phys.{} C~62~(1994)~127\relax
\relax
\bibitem{bookfeynam:1972}
R.P.~Feynman,
\newblock {\em Photon-Hadron Interactions}.
\newblock Benjamin, New York, (1972)\relax
\relax
\bibitem{zfp:c2:237}
K.H. Streng, T.F. Walsh and P.M. Zerwas,
\newblock Z.\ Phys.{} C~2~(1979)~237\relax
\relax
\bibitem{pr:d67:012007}
\colab{ZEUS}, S.~Chekanov \etal,
\newblock Phys.\ Rev.{} D~67~(2003)~012007\relax
\relax
\bibitem{epj:c28:455}
A.D. Martin \etal,
\newblock Eur.\ Phys.\ J.{} C~28~(2003)~455\relax
\relax
\bibitem{jhep:0207:012}
J. Pumplin \etal,
\newblock \JHEP{} 0207~(2002)~012\relax
\relax
\bibitem{jhep:0310:046}
D. Stump \etal,
\newblock \JHEP{} 0310~(2003)~046\relax
\relax
\bibitem{pl:b293:465}
\colab{ZEUS}, M.~Derrick \etal,
\newblock Phys.\ Lett.{} B~293~(1992)~465\relax
\relax
\bibitem{zeus:1993:bluebook}
\colab{ZEUS}, U.~Holm~(ed.),
\newblock {\em The {ZEUS} Detector}.
\newblock Status Report (unpublished), DESY (1993),
\newblock available on
  \texttt{http://www-zeus.desy.de/bluebook/bluebook.html}\relax
\relax
\bibitem{nim:a279:290}
N.~Harnew \etal,
\newblock Nucl.\ Inst.\ Meth.{} A~279~(1989)~290\relax
\relax
\bibitem{npps:b32:181}
B.~Foster \etal,
\newblock Nucl.\ Phys.\ Proc.\ Suppl.{} B~32~(1993)~181\relax
\relax
\bibitem{nim:a338:254}
B.~Foster \etal,
\newblock Nucl.\ Inst.\ Meth.{} A~338~(1994)~254\relax
\relax
\bibitem{nim:a309:77}
M.~Derrick \etal,
\newblock Nucl.\ Inst.\ Meth.{} A~309~(1991)~77\relax
\relax
\bibitem{nim:a309:101}
A.~Andresen \etal,
\newblock Nucl.\ Inst.\ Meth.{} A~309~(1991)~101\relax
\relax
\bibitem{nim:a321:356}
A.~Caldwell \etal,
\newblock Nucl.\ Inst.\ Meth.{} A~321~(1992)~356\relax
\relax
\bibitem{nim:a336:23}
A.~Bernstein \etal,
\newblock Nucl.\ Inst.\ Meth.{} A~336~(1993)~23\relax
\relax
\bibitem{desy-92-066}
J.~Andruszk\'ow \etal,
\newblock Preprint \mbox{DESY-92-066}, DESY, 1992\relax
\relax
\bibitem{zfp:c63:391}
\colab{ZEUS}, M.~Derrick \etal,
\newblock Z.\ Phys.{} C~63~(1994)~391\relax
\relax
\bibitem{acpp:b32:2025}
J.~Andruszk\'ow \etal,
\newblock Acta Phys.\ Pol.{} B~32~(2001)~2025\relax
\relax
\bibitem{nim:a365:508}
H.~Abramowicz, A.~Caldwell and R.~Sinkus,
\newblock Nucl.\ Inst.\ Meth.{} A~365~(1995)~508\relax
\relax
\bibitem{nim:a391:360}
R.~Sinkus and T.~Voss,
\newblock Nucl.\ Inst.\ Meth.{} A~391~(1997)~360\relax
\relax
\bibitem{proc:hera:1991:23}
S.~Bentvelsen, J.~Engelen and P.~Kooijman,
\newblock {\em Proc. of the Workshop on Physics at {HERA}}, W.~Buchm\"uller and
  G.~Ingelman~(eds.), Vol.~1, p.~23.
\newblock Hamburg, Germany, DESY (1992)\relax
\relax
\bibitem{proc:hera:1991:43}
{\em {\rm K.C.~H\"oger}}, ibid., p.~43\relax
\relax
\bibitem{proc:snowmass:1990:134}
J.E. Huth \etal,
\newblock {\em Research Directions for the Decade. Proc. of Summer Study on
  High Energy Physics, 1990}, E.L. Berger~(ed.), p.~134.
\newblock World Scientific (1992).
\newblock Also in preprint \mbox{FERMILAB-CONF-90-249-E}\relax
\relax
\bibitem{tech:cern-dd-ee-84-1}
R.~Brun et al.,
\newblock {\em {\sc geant3}},
\newblock Technical Report CERN-DD/EE/84-1, CERN, 1987\relax
\relax
\bibitem{cpc:69:155}
A. Kwiatkowski, H. Spiesberger and H.-J. M\"ohring,
\newblock Comp.\ Phys.\ Comm.{} 69~(1992)~155\relax
\relax
\bibitem{spi:www:heracles}
H.~Spiesberger,
\newblock {\em An Event Generator for $ep$ Interactions at {HERA} Including
  Radiative Processes (Version 4.6)}, 1996,
\newblock available on \texttt{http://www.desy.de/\til
  hspiesb/heracles.html}\relax
\relax
\bibitem{cpc:81:381}
K. Charchu\l a, G.A. Schuler and H. Spiesberger,
\newblock Comp.\ Phys.\ Comm.{} 81~(1994)~381\relax
\relax
\bibitem{spi:www:djangoh11}
H.~Spiesberger,
\newblock {\em {\sc heracles} and {\sc djangoh}: Event Generation for $ep$
  Interactions at {HERA} Including Radiative Processes}, 1998,
\newblock available on \texttt{http://wwwthep.physik.uni-mainz.de/\til
hspiesb/djangoh/djangoh.html}\relax
\relax
\bibitem{pl:b165:147}
Y. Azimov \etal,
\newblock Phys.\ Lett.{} B~165~(1985)~147\relax
\relax
\bibitem{pl:b175:453}
G. Gustafson,
\newblock Phys.\ Lett.{} B~175~(1986)~453\relax
\relax
\bibitem{np:b306:746}
G. Gustafson and U. Pettersson,
\newblock Nucl.\ Phys.{} B~306~(1988)~746\relax
\relax
\bibitem{zfp:c43:625}
B. Andersson \etal,
\newblock Z.\ Phys.{} C~43~(1989)~625\relax
\relax
\bibitem{cpc:71:15}
L. L\"onnblad,
\newblock Comp.\ Phys.\ Comm.{} 71~(1992)~15\relax
\relax
\bibitem{zfp:c65:285}
L. L\"onnblad,
\newblock Z.\ Phys.{} C~65~(1995)~285\relax
\relax
\bibitem{cpc:101:108}
G. Ingelman, A. Edin and J. Rathsman,
\newblock Comp.\ Phys.\ Comm.{} 101~(1997)~108\relax
\relax
\bibitem{epj:c12:375}
H.L.~Lai \etal,
\newblock Eur.\ Phys.\ J.{} C~12~(2000)~375\relax
\relax
\bibitem{prep:97:31}
B. Andersson \etal,
\newblock Phys.\ Rep.{} 97~(1983)~31\relax
\relax
\bibitem{cpc:82:74}
T. Sj\"ostrand,
\newblock Comp.\ Phys.\ Comm.{} 82~(1994)~74\relax
\relax
\bibitem{cpc:135:238}
T. Sj\"ostrand \etal,
\newblock Comp.\ Phys.\ Comm.{} 135~(2001)~238\relax
\relax
\bibitem{cpc:39:347}
T. Sj\"ostrand,
\newblock Comp.\ Phys.\ Comm.{} 39~(1986)~347\relax
\relax
\bibitem{cpc:43:367}
T. Sj\"ostrand and M. Bengtsson,
\newblock Comp.\ Phys.\ Comm.{} 43~(1987)~367\relax
\relax
\bibitem{np:b485:291}
S. Catani and M.H. Seymour,
\newblock Nucl.\ Phys.{} B~485~(1997)~291.
\newblock Erratum in Nucl.~Phys.~B~510~(1998)~503\relax
\relax
\bibitem{np:b178:421}
R.K. Ellis, D.A. Ross and A.E. Terrano,
\newblock Nucl.\ Phys.{} B~178~(1981)~421\relax
\relax
\bibitem{jp:g26:r27}
S. Bethke,
\newblock J.\ Phys.{} G~26~(2000)~R27.
\newblock Updated in Preprint hep-ex/0606035, 2006\relax
\relax
\bibitem{proc:calor:2002:767}
M. Wing (on behalf of the \colab{ZEUS}),
\newblock {\em Proc. of the 10th International Conference on Calorimetry in
  High Energy Physics}, R. Zhu~(ed.), p.~767.
\newblock Pasadena, USA (2002).
\newblock Also in preprint \mbox{hep-ex/0206036}\relax
\relax
\bibitem{epj:c21:443}
\colab{ZEUS}, S.~Chekanov \etal,
\newblock Eur.\ Phys.\ J.{} C~21~(2001)~443\relax
\relax
\bibitem{jhep:0312:007}
R.W.L. Jones \etal,
\newblock \JHEP{} 0312~(2003)~007\relax
\relax
\bibitem{proc:dis:2005:689}
C. Glasman,
\newblock {\em Proc. of the 13th International Workshop on Deep Inelastic
  Scattering}, S.R. Dasu and W.H. Smith~(eds.), p.~689.
\newblock Madison, USA (2005).
\newblock Also in preprint \mbox{hep-ex/0506035}\relax
\relax
\bibitem{prl:30:1343}
D.J. Gross and F. Wilczek,
\newblock Phys.\ Rev.\ Lett.{} 30~(1973)~1343\relax
\relax
\bibitem{prl:30:1346}
H.D. Politzer,
\newblock Phys.\ Rev.\ Lett.{} 30~(1973)~1346\relax
\relax
\bibitem{pr:d8:3633}
D.J. Gross and F. Wilczek,
\newblock Phys.\ Rev.{} D~8~(1973)~3633\relax
\relax
\bibitem{prep:14:129}
H.D. Politzer,
\newblock Phys.\ Rep.{} 14~(1974)~129\relax
\relax
\bibitem{prl:33:244}
W.E. Caswell,
\newblock Phys.\ Rev.\ Lett.{} 33~(1974)~244\relax
\relax
\bibitem{np:b75:531}
D.R.T. Jones,
\newblock Nucl.\ Phys.{} B~75~(1974)~531\relax
\relax
\bibitem{tmf:41:26}
E.S. Egorian and O.V. Tarasov,
\newblock Theor.\ Mat.\ Fiz.{} 41~(1979)~26\relax
\relax
\end{mcbibliography}
\end{document}